\documentclass[aps,pre,twocolumn,superscriptaddress]{revtex4}

\usepackage[dvips]{graphicx}
\usepackage{amssymb}

\begin{document}
\title{Localized modes and bistable scattering in nonlinear network junctions}

\author{Andrey E. Miroshnichenko$^1$, Mario I. Molina$^2$, and Yuri S. Kivshar}

\affiliation{Nonlinear Physics Centre, Research School of Physical
Sciences and Engineering, Australian National University,
Canberra ACT 0200, Australia\\
$^2$ Departamento de F\'{\i}sica, Facultad de Ciencias, Universidad
de Chile, Santiago, Chile}

\begin{abstract}
We study the properties of junctions created by crossing of $N$
identical branches of linear discrete networks. We reveal that for
$N>2$ such a junction creates a topological defect and supports two
types of spatially localized modes. We analyze the wave scattering
by the junction defect and demonstrate nonzero reflection for any
set of parameters. If the junction  is nonlinear, it is possible to
achieve the maximum transmission for any frequency by tuning the
intensity of the scattering wave. In addition, near the maximum
transmission the system shows the bistable behaviour.
\end{abstract}


\maketitle

\section{Introduction}

Recently, a variety of different discrete network systems have been
studied in order to find the optimal geometries for enhancing both
linear and nonlinear resonant wave
transmission~\cite{chaos,pre_junction,prb_molina1}, to reveal the
conditions of the prefect reflection and Fano
resonances~\cite{prb_molina2,pre_our1,pre_our2}, as well as to
analyze the soliton propagation in the discrete structures where the
role played by the topology of the network becomes
important~\cite{ol_dc,prl_dc,phys_d_junction}. One of the major
issues of those studies is to understand an interplay and
competition of topology and nonlinearity in the dynamics and predict
new interesting phenomena.

Many of such waveguide structures can be described as discrete
linear and nonlinear networks composed of straight, bent, and
crossed waveguides in which forward and backward propagating waves
become coupled to each other via one or more crossings (or network
junctions). The well-known systems for realizing these structures
are photonic-crystal circuits~\cite{prl_fan,pre_minga}, micro-ring
resonator structures with more than two coupled channel
waveguides~\cite{josa_sipe}, and discrete networks for routing and
switching of discrete optical solitons~\cite{prl_dc}. The important
question is how the waveguide crossing affects the wave propagation
in the network and how we can modify the junction transmission by
making it nonlinear.

%
\begin{figure}[htb]
\vspace{20pt} \centerline{
\includegraphics[width=70mm]{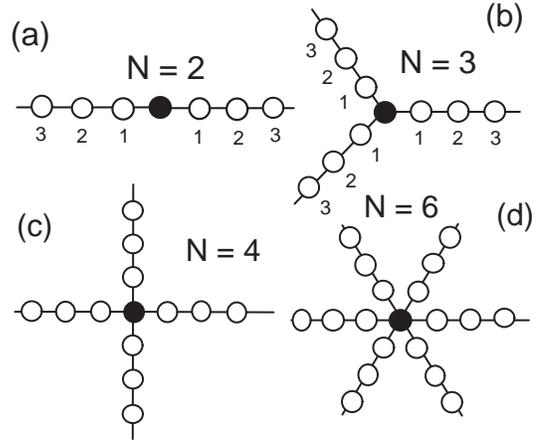}}
\caption{\label{fig1} Examples of $N$-branched discrete network
junctions.
}%
\end{figure}

In this paper, we study the geometry-mediated localized modes and
wave scattering in the structures composed of $N$ discrete
waveguides (`branches') crossing at a common point (`junction'), as
shown schematically in Figs.~\ref{fig1}(a-d). This system can have
direct applications to the physics of two-dimensional
photonic-crystal circuits~\cite{prl_fan,pre_minga}, and it describes
the transmission properties of the $Y$-splitters [see
Fig.~\ref{fig1}(b)], the waveguide cross-talk effects [see
Fig.~\ref{fig1}(c)], and other types of photonic-crystal devices. We
show that the intersection point of the discrete waveguides acts as
a complex $\delta$-like defect even in the absence of any site
energy mismatch, i.e. when the crossing point is identical to any of
the waveguide points. Therefore, the waveguide junction can be
viewed as \textit{a topological defect}. We show that this new type
of defect gives rise to two stable localized modes (or defect modes)
at the intersection point, the staggered and unstaggered ones,
regardless of the amount of nonlinearity or site energy mismatch at
the junction. Next, we study the scattering of plane waves by the
crossing point, where the incoming wave propagates along one of the
branches, and the waves are scattered to all the branches. In the
linear regime, we find that for $N>2$ the reflection coefficient is
always nonvanishing. This fundamental limitation prevents us from
achieving an ideal $Y$-splitter, unless we optimize and engineer the
sites near the junction~\cite{ol_dc}. For a nonlinear junction of
the crossed discrete linear waveguides, we reveal the possibility of
achieving the maximum transmission for almost any frequency of the
incoming plane wave by tuning the input intensity. For some
conditions, bistability in the transmission may occur near the
transmission maximum. To verify our analytical results obtained for
plane waves, we study the propagation of a Gaussian pulse across the
junction by direct numerical simulations. The numerical results
agree nicely with our analysis.

The paper is organized as follows. In Sec.~II we introduce our
model. Section~III is devoted to the study of the properties of a
linear junction, while Sec.~IV considers the case of a nonlinear
junction and discuss bistability. Our numerical results are
summarized in Sec.~V, while Sec.~VI concludes the paper.

\section{Model}

We consider the system created by $N$ identical discrete linear
waveguide (branches) crossed at a common point (junction) which we
consider to be linear or nonlinear [see Figs.~\ref{fig1}(a-d)]. The
corresponding waveguide structure can be described in a general form
by the system of coupled discrete nonlinear equations,
\begin{eqnarray}
\label{eq1}
i\dot{\phi}_n^{(k)}&=&\phi_{n+1}^{(k)}+\phi_{n-1}^{(k)}\;,\;n\ge1\nonumber\\
\phi_0^{(k)}&=&\phi_0\;,\\
i\dot{\phi}_0&=&\sum\limits_{k=1}^N\phi_1^{(k)}+ (\epsilon+\lambda|\phi_0|^2)\phi_0\;,\nonumber
\end{eqnarray}
where the index $k$ refers to the branch number ($k=1,2,\cdots, N$).
In this model, we endow the junction ($n=0$) with a linear impurity
$\epsilon_0$ and a Kerr nonlinear coefficient $\lambda$. We notice
that a similar discrete model can be derived for different types of
two-dimensional photonic-crystal devices based on the Green's
function approach~\cite{pre_minga}. In that case, the complex field
$\phi_n^{(k)}$ corresponds to the amplitude of the electric field.

\section{Linear junction}

First, we consider the linear regime when $\lambda=0$, and all
equations (\ref{eq1}) are linear. We start our analysis by
investigating the possible localized states in the system
(\ref{eq1}). In order to do that, we look the solutions in the
well-known form,
\begin{eqnarray}
\label{eq2} \phi_n^{(k)}&=&A
\exp\{-\chi|n|-i\Omega\tau\}\;,\;n\ge0\;,
\end{eqnarray}
where $A$ is the mode amplitude, $\tau$ is the corresponding
evolution coordinate (time in a general case, or the longitudinal
propagation coordinate, for some problems of optics), $\Omega$ is
the frequency, and $\chi$ is the spatial (or temporal) decay rate of
a localized state. After substituting this solution into
Eq.~(\ref{eq1}), we obtain the relations for the frequency $\Omega$
and the decay rate $\chi$,
\begin{eqnarray}
\label{eq3}
\Omega&=&2\cosh\chi\;,\nonumber\\
e^{\chi}&=&(N-1)e^{-\chi}+\epsilon_0\;.
\end{eqnarray}
By denoting $y=e^{-\chi}$, we rewrite the second equation in
Eq.~(\ref{eq3}) as follows
\begin{eqnarray}
\label{eq4}
(N-1)y^2+\epsilon_0y-1=0\;,
\end{eqnarray}
which, in general, possesses two solutions
\begin{eqnarray}
\label{eq5} y^{\pm}={1\over{N-1}}\left(
-\frac{\epsilon_0}{2}\pm\sqrt{\left(\frac{\epsilon_0}{2}\right)^{2}+N-1}
\right),\label{eq:5}
\end{eqnarray}
which are of opposite signs $y^{+}>0, y^{-}<0$. As a result, there
exist two different solutions of our system
\begin{eqnarray}
\label{eq6}
\chi^{\pm}=\left\lbrace%
    \begin{array}{ll}
     -\ln y^{+}\\
     -\ln|y^{-}|+i\pi,
    \end{array}
 \right.
\end{eqnarray}
which describe unstaggered or staggered localized states.

%
\begin{figure}[htb]
\vspace{20pt} \centerline{
\includegraphics[width=70mm]{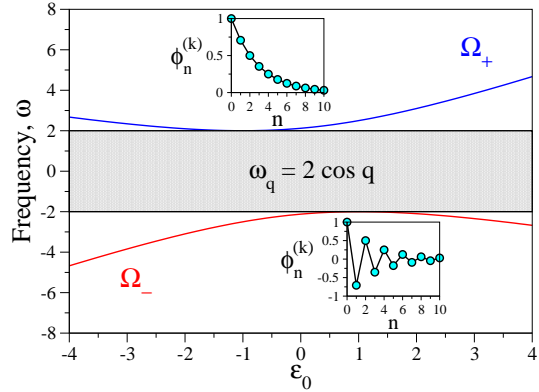}}
\caption{\label{fig2}(Color online) Spectrum of linear waves of the system
(\ref{eq1}) vs. the impurity strength $\epsilon_0$ for $N=3$. In
addition to the propagating modes $\omega_{q}=2\cos q$, two
localized states appear, $\Omega^{\pm}$. At the critical values
these localized states touch the band of the propagating modes.
Insets show the amplitude profiles of the corresponding staggered
and unstaggered modes for $\epsilon_0=0$.
}%
\end{figure}

In the simplest case of a homogeneous system ($\epsilon_0=0$) there
exist always two bound states for $N\ge2$ [see Fig.~\ref{fig2}],
\begin{eqnarray}
\label{eq7}
\chi^{+}=\ln(\sqrt{N-1})\;\;,\;\;\chi^{-}=\ln(\sqrt{N-1})+i\pi\;.
\end{eqnarray}
Therefore, the junction itself constitutes an example of a different
class of defect: \textit{a topological defect}. This defect supports
two localized states whose decay rates $\chi^{\pm}$ are inversely
proportional to the number of branches $N$.

In an inhomogeneous case when $\epsilon_0\not=0$, the decay rate of
the localized state can change drastically depending on the impurity
strength. For example, Eq.~(\ref{eq4}) supports the solutions $y=1$
(or $\chi=0$), when $\epsilon_0=2-N$ and $y=-1$ (or $\chi=i\pi$),
when $\epsilon_0=N-2$. This means that in these cases one of the
localized states can disappear. By looking at the frequency
dependence of the localized state $\Omega$ on the impurity strength
$\epsilon_0$ see Fig.~\ref{fig2}], we see that one of the defect
modes touches the band of the linear spectrum exactly when the
corresponding localized state disappears. This happens precisely at
$\epsilon_{0} = \pm (N-2)$.

Now we analyze the transmission properties of this new type of
defect. To this end, we consider an incoming plane wave propagating
in the branch $k=1$ and calculate the transmitted waves in other
branches. We begin by imposing the scattering boundary conditions
for Eq.~(\ref{eq1}),
\begin{eqnarray}
\label{eq8}
\phi_n^{(k)}=e^{-i\omega_q\tau}\left\lbrace%
    \begin{array}{lc}
     Ie^{iqn}+re^{-iqn}& n>1,\; k=1\;,\\
     te^{iqn}& n>1,\; k>1\;,
    \end{array}
 \right.
\end{eqnarray}
where $\omega_q=2\cos q$ is the the frequency of the incoming plane
wave. Continuity condition of the plane wave at the junction implies
$I+r=t$. On the other hand, the equation for $\phi_0$ can be written
in the form,
\begin{eqnarray}
\label{eq9}
\omega_qt=IE^{-iq}+re^{iq}+(N-1)te^{iq}+\epsilon_0t\;.
\end{eqnarray}
%
%
\begin{figure}[htb]
\vspace{20pt} \centerline{
\includegraphics[width=70mm]{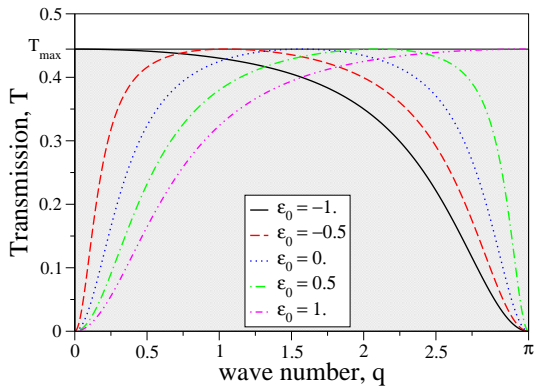}}
\caption{\label{fig3}(Color online) Linear transmission coefficient
for different values of the impurity strength $\epsilon_0$ (and
$N=3$). This plot shows that at the critical values
$\epsilon_0=\pm(2-N)$, the transmission coefficient becomes maximum
at the corresponding band edge, $T(0,\pi)=4/N^2$, when one of the
localized states touches the edge of the linear spectrum [see
Fig.~\ref{fig2}]. For intermediate values, $2-N<\epsilon_0<N-2$, the
maximum transmission is inside the propagation band. The shaded area
indicates the possible values of the transmission coefficient.
}%
\end{figure}

By combining these two conditions, we can obtain the amplitude of
the transmitted wave,
\begin{eqnarray}
\label{eq10} t=\frac{2i\sin qI}{2i\sin
q+\epsilon_0+(N-2)e^{iq}}=\frac{2}{N}\frac{I}{1-i\alpha_q}\;,
\end{eqnarray}
where $\alpha_q=(\epsilon_0+(N-2)\cos q)/(N\sin q)$. The
transmission coefficient for any branch is defined as $T=|t/I|^2$.
From Eq.~(\ref{eq10}) we see that even in the homogeneous system
($\epsilon_0=0$), the waveguide junction produces an effective
geometry-induced complex $\delta$-like scattering potential,
$\epsilon_q=(N-2)e^{iq}$, which depends on the parameters of the
incoming plane wave.

We can write the transmission coefficient in the form
\begin{eqnarray}
\label{eq11} T=\frac{4}{N^2}\frac{1}{1+\alpha_q^2}\;,
\end{eqnarray}
the reflection coefficient is: $R=1-(N-1)T$. The maximum of the
transmission coefficient $T_{\rm max}=4/N^2$ can be reached for
$\alpha_q=0$. We note that $R_{\rm min}=1-4(N-1)/N^2>0$
for $N>2$, and it never vanishes. As a result, the $N$-junction
system described by the model (\ref{eq1}) will always reflect some
energy back into the incoming branch. This is a fundamental
limitation which does not allow us to build a perfect $Y$-splitter.

At this point, the following natural question arises: {\em What
happens in the system at the maximum transmission?} First, we notice
that this point does not correspond to any resonance. According to
the linear scattering theory all (quasi-)bound states of our system
can be obtained as poles of the the transmission amplitude
(\ref{eq10}). The poles can be found in the complex plane from the
condition $\alpha_q=-i$, by assuming that the wave number $q$ is
complex $q=q^r+i\chi$. Complex wave numbers correspond to
quasi-bound states with the complex frequency $\Omega$, and they can
be interpreted as resonances where the real part of the frequency
gives the resonance frequency and its imaginary part describes the
lifetime (or width) of the resonance. After some algebra, we find
that there exist only two solutions for $q^r=0,\pi$, and for $\chi$
we have exactly the same solutions as in Eq.~(\ref{eq4}). This means
that our system does not support any quasi-bound states
($0<q^r<\pi$), and there exist only two real bound states
($q^r=0,\pi$, $\chi\not=0$).

What really happens is quite the opposite. Due to the frequency
dependence of the effective scattering potential $\epsilon_q$, it
simply disappears when the condition $\epsilon_0+Re(\epsilon_q)=0$
(or $\alpha_q=0$) is satisfied. And the system becomes almost
transparent. The nonzero reflection exists due to the nonzero
imaginary part of the effective scattering potential
$Im(\epsilon_q)=(N-2)\sin q$.

Equation for poles of the transmission amplitude $\alpha_q=-i$
differs from the equation for the maximum transmission $\alpha_q=0$,
and there is no relation between them. But, in fact, there exists
the condition when these two equations may produce the same results,
e.g. when $q=0,\pi$ and one of the bound state disappears. In that
case we observe the maximum transmission at the corresponding band
edge. This result can be understood in terms of Levinson theorem,
where a bound state just enters to or emerges from the propagation
band and forms a quasi-bound state.

Now we can compare the dependencies of the defect-mode frequencies
$\Omega^{\pm}=2\cosh\chi^{\pm}$ and the transmission coefficient $T$
on the impurity strength $\epsilon_0$ [see Fig.~\ref{fig3}]. When
$\epsilon_0=2-N$, the first unstaggered localized state $\Omega^{+}$
touches the upper band edge, and the maximum transmission takes
place at that edge. For intermediate values of the impurity
strength, $2-N<\epsilon_0<N-2$, we observe the maximum transmission
inside the propagation band which moves towards another edge.
Finally, when $\epsilon_0=N-2$ the second staggered localized state
$\Omega^{-}$ touches the bottom edge together with the occurrence of
maximum transmission.

\section{Nonlinear junction}

Now we consider the nonlinear junction and analyze both the
localized states and wave transmission. For $\lambda\not=0$, the
junction can support nonlinear localized states (\ref{eq2}). The
decay rates $\chi^{\pm}$ can be found from the equation similar to
Eq.~(\ref{eq4}),
\begin{eqnarray}
\label{eq12}
(N-1)y^2+\tilde{\epsilon_0}y-1=0\;,
\end{eqnarray}
where we have renormalized the impurity strength,
$\tilde{\epsilon_0}=\epsilon_0+\lambda|A|^2$. Thus, all previous
results about linear localized states remain qualitatively valid in
the nonlinear regime. The only difference is that in the latter
case, there is an additional dependence on the intensity of the
localized state, $|A|^2$. For instance, when nonlinearity is
focussing $\lambda>0$, we have $\tilde{\epsilon_{0}}>\epsilon_{0}$.
From Eq.~(\ref{eq:5}) it follows that $y^{+}$ decreases while
$y^{-}$ increases. This, in turn, implies that the corresponding
modes $\Omega^{+}$ and $\Omega^{-}$ become narrower and broader,
respectively. In the limit of large nonlinearity, the staggered
localized mode will extend all over the branches, while the
unstaggered mode will remain confined to essentially one site of the
junction. Stability analysis shows, that in both cases, the
localized modes remain stable. We find no other localized modes,
even in the strong nonlinearity regime.

%
%
\begin{figure}[htb]
\vspace{20pt} \centerline{
\includegraphics[width=70mm]{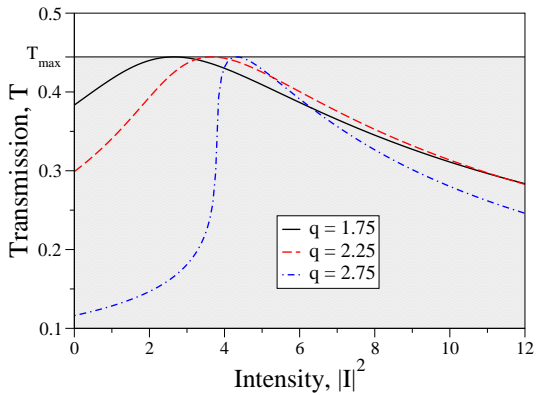}}
\caption{\label{fig4}(Color online) Nonlinear transmission vs.
intensity of the incoming wave $|I|^2$ for different values of the
wave number $q$. Other parameters are: $N=3$, $\epsilon_0=-1$, and
$\lambda=1$. The maximum transmission can be achieved for any value
of the wave number $q$ by tuning the intensity $|I|^2$ of the
incoming wave. The shaded area indicates the possible values of the
transmission coefficient.}
\end{figure}
%

\begin{figure}[htb]
\vspace{20pt} \centerline{
\includegraphics[width=70mm]{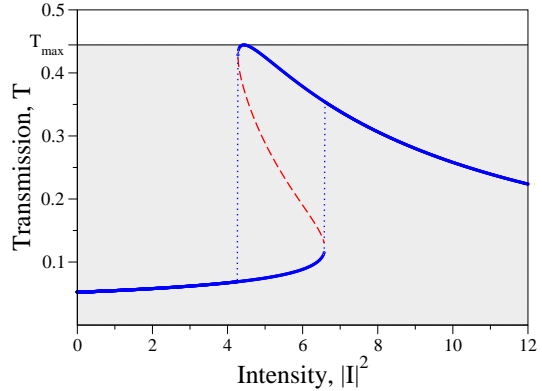}}
\caption{\label{fig5}(Color online) Bistable nonlinear transmission
in the nonlinear junction for $q=2.9$. Other parameters are the same
as in Fig.~\ref{fig4}. Bistability takes place near the maximum
transmission. The shaded area indicates the possible values of the
transmission coefficient.
}%
\end{figure}

\begin{figure}[b]
\vspace{20pt} \centerline{
\includegraphics[width=70mm]{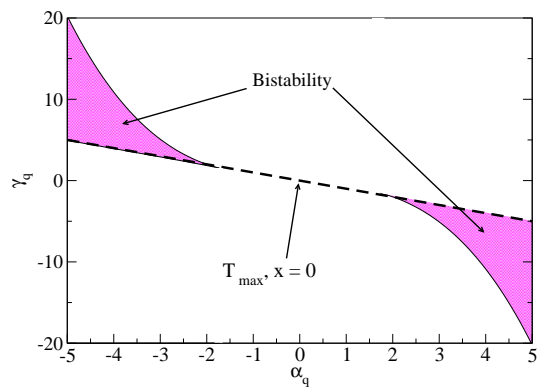}}
\caption{\label{fig6}(Color online) Bistability diagram described by
the solutions of the cubic equation (\ref{eq14}). For the system
parameters inside the shaded region there exist three solutions of
the cubic equation, and the bistable nonlinear transmission should
be observed (see Fig.~\ref{fig5}). The solution $x=0$ corresponds to
the maximum transmission $T_{\rm max}=4/N^2$, and it is shown as a
dashed line $\gamma_{q} = -\alpha_{q}$.
}%
\end{figure}

The presence of nonlinear junction generates much richer dynamics for
the transmission. The transmission coefficient can now be written in
the form,
\begin{eqnarray}
\label{eq13} T=\frac{4}{N^2}\frac{1}{(x^2+1)}\;,
\end{eqnarray}
where $x$ satisfies the cubic equation
\begin{eqnarray}
\label{eq14}
(x^2+1)(x-\alpha_q)-\gamma_q =0,
\end{eqnarray}
$\gamma_q=\lambda 4I^2/(N^3\sin q)$ is the normalized nonlinear
parameter, and $c_q=2\sin q$. The maximum transmission $T_{\rm
max}=4/N^2$ takes place when $x=0$ is a solution of
Eq.~(\ref{eq14}), i.e.,  when $\alpha_q=-\gamma_q$. It implies that
the maximum transmission can be achieved for any frequency by a
proper tuning of the intensity of the incoming wave
[Fig.~\ref{fig4}]. Moreover, the analysis of the cubic equation
(\ref{eq14}) reveals that for $|\alpha_q|>\sqrt{3}$ three solutions
are possible, and the bistable transmission should occur
(Fig.\ref{fig5}). We summarize all those scenarios in
Fig.~\ref{fig6}. We notice that the maximum transmission curve,
$\gamma_{q} = -\alpha_{q}$, lies almost at the boundary of the
bistability region.

%

\section{Results of numerical simulations}

%

In order to verify our theoretical results, we perform direct
numerical simulations of the system (\ref{eq1}) under more realistic
pulse propagation. We launch a Gaussian pulse, along the branch
$k=1$ and study numerically its propagation through the junction,
\begin{eqnarray}
\label{eq15} \phi_n^{(1)}(0) =
I_0\exp\{-\frac{(n-n_0)^2}{\sigma^2}-iq_0(n-n_0)\}\;,
\end{eqnarray}
where $q_0$ is the pulse momentum, $I$ is the maximum amplitude of
the wavepacket, $\sigma$ is the spatial width, and $n_0$ is the
initial position.

%
\begin{figure}[htb]
\vspace{20pt} \centerline{
\includegraphics[width=70mm]{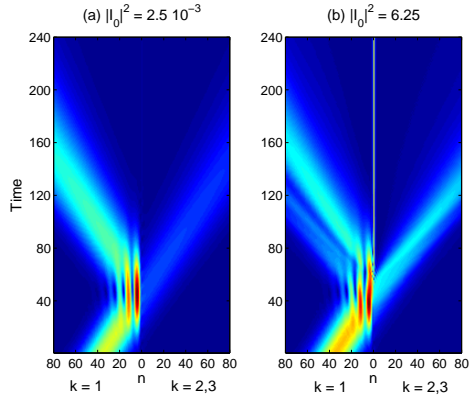}}
\caption{\label{fig7}(Color online) Evolution of a Gaussian pulse
scattered by the $N=3$ nonlinear junction (a nonlinear
$Y$-splitter), for two different pulse intensities: (a)
$|I|^2=2.5\times10^{-3}$, and (b) $I=6.25$. Other parameters are:
$\epsilon_0=-1$, $\lambda=1$, $\sigma=20$ and $q_0=2.75$. These
results should be compared with the results for the plane-wave
scattering shown in Fig.~\ref{fig4}.
}%
\end{figure}

For our numerical simulations, we consider the case of a nonlinear
$Y$-junction splitter ($N=3$) with $\epsilon_0=-1.$ and $\lambda=1.$
[see Fig.~\ref{fig4}]. Figure~\ref{fig7} shows our results for the
wave number $q_0=2.75$ for two different intensities of the Gaussian
pulse. In this figure, we show the temporal pulse evolution in two
branches only ($k=1$ and $k=2$) since, because of symmetry, the
evolution in the branches $k=2$ and $k=3$ coincide.

because the evolution in the third branch $k=3$ coincides with that
in the branch $k=2$ due to a symmetry.

When the pulse intensity  is small [see Fig.~\ref{fig7}(a)], the
whole Gaussian pulse is reflected back into the incoming branch
$k=1$, in agreement with our theoretical results [Fig.~\ref{fig4}].
For larger values of the pulse intensity [see Fig.~\ref{fig7}(b)],
we observe an almost optimal splitting of the Gaussian pulse into
the branches $k=2$ and $k=3$ with the maximum transmission [see
Fig.~\ref{fig4}]. We notice that in this case the junction $\phi_0$
remains highly excited even after the pulse already passed through
it. The excitation of the nonlinear localized state at the junction
is possible because of $\epsilon_{0}=-1$, one of the localized
states interact strongly with the linear spectrum band (and its
eigenvalue touches the band edge, see Fig.~\ref{fig2}).

\section{Conclusions}

We have analyzed the linear and nonlinear transmission through a
junction created by crossing of $N$ identical branches of a network
of discrete linear waveguides. We have revealed that for $N>2$ such
a junction behaves as an effective topological defect and supports
two types of spatially localized linear modes. We have studied the
transmission properties of this junction defect and demonstrated
analytically that the reflection coefficient of the junction never
vanishes, i.e. the wave scattering is always accompanied by some
reflection. We have considered the case when the junction defect is
nonlinear, and studied the nonlinear transmission of such a local
nonlinear defect. We have demonstrated that nonlinearity allows
achieving the maximum transmission for any frequency by tuning the
intensity of the incoming wave but, in addition, the system can
demonstrate bistability near the maximum transmission. We have
confirmed our analytical results by direct numerical simulations.

\section*{Acknowledgments}

This work has been supported by the Australian Research Council in
Australia, and by Fondecyt grants 1050193 and 7050173 in Chile.

\end{document}